\documentclass{article}
\usepackage{graphicx} 

\title{Courses}
\author{s.mahdavi.hezavehi }

\usepackage{natbib}
\usepackage{url}

\title {State of the Art on Self-adaptive Systems: \\An Essay}
\author{Sara Mahdavi Hezavehi \\Supervisors: Danny Weyns and Paris Avgeriou}

\begin{document}

\maketitle
\section{Self-adaptive Systems - An Overview}
In this essay, we introduce the basic concepts necessary to lay out the foundation for our PhD research on uncertainty and risk-aware adaptation, and discuss relevant related research.  

\subsection{Self-Adaptive System}
In this chapter, we consider a system to be self-adaptive if it complies with two basic principles~\cite{weyns2020book}. The \textit{external principle} states that a self-adaptive system can achieve its goals in the face of changes and uncertainties without or with minimal human intervention. While a human operator manages uncertainties in a regular computing system, a self-adaptive system can deal with uncertainties autonomously, possibly supported by an operator, taking a set of adaptation goals as input. 
The \textit{internal principle} states that a self-adaptive system comprises two distinct parts: the first part, called the \textit{managed system}, interacts with the environment and is responsible for the domain concerns for which the system is built; the second part, called the \textit{managing system}, consists of a feedback loop that interacts with the first part (and monitors its environment) and is responsible for the adaptation concerns, i.e., concerns about the domain concerns. For instance, a domain concern for an IoT system may be to collect data obtained by its nodes deployed in the environment, and an adaptation concern may be to collect this data with minimum energy consumption. A human operator may be involved in realizing the functions of the managing system (human-in-the-loop), or in observing the system in operation and only taking action when needed (human-on-the-loop).

\subsection{Architecture-based Adaptation}
Our particular focus is on \textit{architecture-based adaptation}~\cite{Oreizy1999,Garlan2004,Kramer2007,Weyns2012}, which is an established approach to realize self-adaptive systems. Architecture-based adaptation has a dual focus~\cite{weyns2020book}: on the use of software architecture as an abstraction to \textit{define} a self-adaptive system at design time~\cite{Kramer2007}; and on the use of architectural models to \textit{reason} about change and make adaptation decisions at runtime~\cite{Garlan2004}. We are primarily concerned with the second aspect. 

Aligned with the second principle of self-adaptation, architecture-based adaptation comprises a managed system (e.g., an IoT system with battery-powered nodes that are deployed in the environment and communicate wireless)   that deals with the domain concerns, i.e., the functions or services that need to be provided to the users (e.g., the network collects data about the presence of humans, the status of locks, the temperature, etc.) , and a managing system that deals with the adaptation concerns, i.e., how the domain concerns are achieved in terms of benefits (e.g., allowed packet loss, minimum energy consumption), costs (e.g., the extra energy required to communicate adaptation actions to the nodes in order to change the network settings), and risks (e.g., the level of privacy protection of the data communicated over the IoT network).  
A reference approach to realize the managing system is by means of a so-called MAPE feedback loop~\cite{Kephart2003,6595487}. MAPE refers to the basic functions that need to be realized by the feedback loop: Monitor the system and its environment, Analyze the situation and the options for adaptation, Plan the adaptation of the managed system for the best adaptation option, and Execute the actions of the plan to adapt the managed system. The MAPE functions share a repository with Knowledge that typically comprises different types of runtime models~\cite{2009Blair,Weyns2012} (MAPE is therefore also referred to as MAPE-K). A concrete architecture maps the MAPE-K functions to components, which can be one-to-one or any other mapping (e.g., analysis and planning may be mapped to a decision-making component). 

In 1998, \cite{Oreizy1999} presented a pioneering model for architecture-based adaptation that comprises two simultaneous processes: system adaptation that deals with detecting and handling changes, and system evolution that deals with the consistent application of change over time. Around the same time, IBM launched its legendary initiative on autonomic computing~\cite{1160055} that took inspiration from the autonomic nervous system of the human body to enable computing systems to manage themselves based on high-level goals. \cite{Garlan2004} pointed out the central role of architectural models as first-class elements that enable a system to reason about system-wide change,  
a principle that aligns with ``models at runtime'' introduced by \cite{2009Blair}. Since these pioneering efforts, a substantial body of knowledge has been developed in this field~\cite{weyns2020book}; some characteristic examples include~\cite{Calinescu2011,10.1145-2593929.2593944,10.1145-2465478.2465479}. A recent large-scale survey~\cite{10.1145-3589227} provided evidence that the principles of architecture-based adaptation are widely applied in industry.

\subsection{Uncertainty in Self-Adaptive Systems}
Already in 2010, \cite{Garlan2010} pointed to the key role of uncertainty in software-intensive systems. 
Different researchers have provided taxonomies for uncertainty in self-adaptive systems, including  \cite{Ramirez2012}, \cite{icpe14}, \cite{Esfahani2013}, and \cite{Musil2017}. Leveraging these efforts, \cite{Mahdavi2017} provided a systematic overview of uncertainty dimensions (location, nature, level/spectrum, emerging time, sources) with their respective options. In this overview, the sources of uncertainty are further elaborated and are grouped into several classes, i.e., uncertainty of models, adaptation functions, goals, environment, resources, and managed system. 

Recently, \cite{Survey2021} identified existing notations and formalisms used to represent the different types of uncertainty, together with the software development phase in which they are used and the types of analysis allowed. \cite{10.1145-3487921} performed a survey in which over 100 members of the research community provided insights into how the concept of uncertainty is understood and currently handled in the engineering of self-adaptive systems. 

According to \cite{weyns2020book}, the fifth of seven waves of research in engineering self-adaptive systems focuses on ``Guarantees Under Uncertainties'' emphasizing uncertainty as a core driver for self-adaptation. In consolidating the existing work on self-adaptive systems, \cite{weyns2020book} defines uncertainty in self-adaptive systems as ``any deviation of deterministic knowledge that may reduce the confidence of adaptation decisions made based on the knowledge.'' The research presented in our work aligns with this definition. To that end, we focus on mitigating uncertainty in the decision-making of self-adaptive systems by taking into account benefit, cost, and risk when an adaptation option is considered and ultimately selected for execution.

\subsection{Decision-Making in Self-Adaptive Systems based on Estimated Benefit}
Most existing approaches for runtime decision-making in self-adaptive systems focus on the benefit that can be obtained when applying an adaptation to the managed system. 
Estimated benefit refers to the expected advantage (or implied effects) that will be obtained by adapting the system from its current configuration to a new configuration. Estimated benefit is usually expressed in terms of quality properties of the system in the form of adaptation goals~\cite{6224395}. 

We summarize several prototypical approaches for decision-making in self-adaptive systems that are based on the estimated benefit. \cite{MorenoCGS15} proposed a method for improving decision–making in a self-adaptive system that is based on maximizing the accumulated utility over the look-ahead horizon. In the example case of a server-based system used for evaluation, the utility was defined as a weighted sum of the average time to serve a request and the cost to be paid for the adaptation options. ~\cite{10.5555/2050655.2050707} presented a goal-based requirements model-driven approach for automatically deriving state-, metric-, and fuzzy logic-based utility functions for relaxed goal models. This approach was evaluated for an intelligent vehicle system. \cite{CAMARA201851} presented a formal approach based on a stochastic game that considers uncertainty in sensing when reasoning about the best way to adapt and improve system utility. Utility in a client-server case used for evaluation is defined based on user annoyance and the portion of malicious clients obtained when selecting a configuration for adaptation. \cite{8008800} presented ENTRUST, a framework that combines (1) design-time and runtime modeling and verification, with (2) industry-adopted assurance cases. The approach employs probabilistic verification to verify stochastic models of the adaptation options of the managed system and its environment to comply with a set of rules. In the case of an unmanned underwater vehicle used for evaluation, the rules included a minimum threshold for measurement accuracy and minimizing sensing energy.  \cite{WeynsI0M18} verified adaptation options using statistical model checking to ensure that the selected option complies with a set of adaptation goals that are defined as rules. In the example of an Internet-of-Things case used for the evaluation, rules included thresholds on allowed packet loss and network latency, and minimization of the energy consumed for communication. \cite{Purandare23} applied adaptation with three strategies to ensure the safety of small Uncrewed Aerial Systems (sUAS): stabilize the sUAS so that they can complete their flight, send the sUAS into loiter mode, and land the sUAS in place. An algorithm evaluates large sets of data to determine the stability of the system and its current context and based on that applies one of the strategies if needed. Other examples on obtaining benefits from applying self-adaptation are \cite{DBLP:conf/ssbse/BowersFC18,Bencomo2013}. For the vast body of work on decision-making in self-adaptation based on estimated benefit, we refer to the literature. 

In summary, while each of these approaches provides valuable contributions to the research on self-adaptive systems, they all consider only the estimated benefit of adaptation options when selecting a new configuration for adapting the managed system.

\subsection{Decision-Making in Self-Adaptive Systems based on Estimated Benefit and Cost}

A number of self-adaptation approaches also consider the estimated cost, in addition to the estimated benefit, when making adaptation decisions. The estimated cost refers to the expected (one-off) cost of executing the option selected for adaptation. Merely focusing on estimated benefit and disregarding the impact of the cost implied by adapting the system may adversely affect the expected benefit of adaptation, and hence the quality goals of the system~\cite{VanDerDonckt2018}. Cost may refer to time or resources that are required to apply adaptation options. Resources may refer to CPU cycles, network bandwidth, disc, memory, and battery energy to provide services to the user.  Yet, there is currently no clear view in the literature on different types or classes of estimated cost; cost is often domain-specific and varies from system to system. We highlight a number of representative approaches.  

\cite{10.1145/3184407.3184413} point to costs for self-adaptation, such as planning delay and extra resource/energy consumption. To that end, they address the problem of how to make a binary decision at each point in time: whether to adapt the self-adaptive system, considering the dynamic and uncertain monetary cost-benefit of adapting the system or not. The approach leverages principles from technical debt and online machine learning. \cite{bertolli} presented an approach that dynamically selects components according to an adaptation strategy to achieve a required level of quality of service. The approach includes both a performance model and a cost model that quantifies the overhead for reconfiguring a component (e.g., when switching between versions). The approach is evaluated on a flood management application. 
\cite{10.1145/2774222} considered cost in terms of adaptation tactic latency, i.e., the interval between the time when a tactic’s execution is triggered and the time when its effects are observed in the state of the system due to application of adaptation. Similarly, \cite{5069082} considered cost as the time difference between enacting an adaptation tactic and the time when its effects can be observed. 
\cite{10.1007/978-3-030-22559-9_17,10.5220/0006815404780490} considered cost as a first-class concern in selecting adaptation options at runtime. Their approach leverages the Cost-Benefit-Analysis Method \cite{CBAMWebsite}, transferring this classic architecture selection method to runtime. The approach is applied to an IoT application where the cost corresponds to the energy required to send adaptation actions from the managed system deployed at the gateway to the nodes of the IoT network. 

In summary, several approaches have demonstrated the value of considering the cost of applying adaptation actions. Their main focus is on the expected time between enacting an adaptation action and observing its effects, and on the resources required to apply the adaptation actions.

\subsection{Estimated Risk in the Decision-making of Self-adaptive Systems}
Estimated risk is rarely considered in the decision-making process of self-adaptive systems. With estimated risk, we refer to potential effects of uncertainties on
system objectives in terms of their likelihoods and consequences (positive, negative, or both)~\cite{RiskWebsite}. Traditionally, risk estimation in software is a human-driven activity performed by system designers and architects at design time. However, in self-adaptive systems, the risk may change over time, possibly impacting users as well as the system itself. To address this issue, we argue that the risk should be addressed throughout the entire life cycle of the system. This implies that in addition to cost and benefit, risk estimation should be made an indispensable concern of runtime decision-making in self-adaptive systems. 

 \cite{6214772} apply a risk-based approach to define change loads for resilience benchmarks for self-adaptive systems. In \cite{6214772}, Reichstaller and Knapp target testing of self-adaptive systems, focusing on risk-based goals and the detection of hazardous failures. One approach that refers to risk in self-adaptive systems is described by \cite{7968127}. The authors highlight that the
satisfaction rate of the goals of self-adaptive systems depends on the rate at which adverse conditions prevent their satisfaction. Obstacle analysis is a goal-oriented approach to risk analysis where obstacles to system goals are identified, assessed, and resolved through countermeasures, yielding new goals. The selection of appropriate countermeasures relies on the assessed likelihood and criticality of obstacles together with environmental assumptions. To meet the system’s goals under changing conditions, the authors proposed an approach for runtime obstacle resolution. The approach relies on a model where goals and obstacles are refined and specified in a probabilistic linear temporal logic. The approach allows for (a) monitoring the satisfaction rate of probabilistic leaf obstacles; (b) determining the severity of their consequences by up-propagating satisfaction rates through refinement trees from leaf obstacles to high-level probabilistic goals; and (c) dynamically shifting to alternative countermeasures that better meet the required satisfaction rate of the system’s high-level goals under imposed cost constraints.
The approach is evaluated for an ambulance dispatching application.

In summary, estimated risk in the decision-making process of self-adaptive systems has been largely ignored so far. Given the growing importance of risk mitigation in terms of safety/privacy of users, environmental impact, and ethical or legal concerns, it is crucial to incorporate estimated risk as a first-class concern in the decision-making of self-adaptive systems. 

\subsection{Reference Model}

There is no common definition of what constitutes a reference model. In this chapter, we refer to a reference model as a set of functional entities with relationships between these entities that together solve a given problem. A reference model is abstractly defined and is domain and technology-agnostic. A concrete architecture maps the functions to concrete components. 

The pioneering reference model in the field of self-adaptation is MAPE-K~\cite{Kephart2003}. MAPE-K defines the essential functions of a feedback loop of a self-adaptive system and their relationships. We highlight two other well-known reference models for self-adaptation. First, FORMS, short for FOrmal Reference Model for Self-adaptation~\cite{Weyns2012}, provides a small set of formally specified modeling elements that correspond to the key concerns in the design of self-adaptive software systems, and a set of relationships that guide their composition. FORMS provides three complementary perspectives: computational reflection, distributed coordination, and MAPE-K. Second, DYNAMICO, short for Dynamic Adaptive, Monitoring and Control Objectives model ~\cite{VillegasTMDC10}, aims at addressing: (i) the management of adaptation properties and goals as control objectives; (ii) the separation of concerns among feedback loops required to address control objectives over time; and (iii) the management of dynamic context as an independent control function to preserve context-awareness in the adaptation mechanism. To that end, the DYNAMICO reference model integrates three types of feedback loops that focus on the control objectives, the target system adaptation, and dynamic monitoring, respectively.

In summary, the field of self-adaptation has produced a number of reference models that define core functions of self-adaptive systems. By consolidating existing knowledge in the decision-making of self-adaptive systems, our work aims to outline a reference model for decision-making in self-adaptive systems that takes into account benefit, cost, and risk as core concerns. 

\subsection{Architectural Viewpoint and View}

Architectural viewpoints and views are a common approach to documenting the architecture of software-intensive systems~\cite{HOFMEISTER2007106,4278472}. The IEEE 1471 and ISO/IEC 42010 standards ~\cite{4278472} offer widely accepted conceptual definitions of architectural viewpoints and views. Specifically, an \textit{architectural viewpoint} is defined as ``a work product establishing the conventions for the construction, interpretation, and use of architecture views to frame specific system concerns.'' 
An architectural viewpoint gives architects the means to express a coherent set of concerns, the stakeholders interested in these concerns, and model kinds (i.e., meta-models) that frame the concerns, each defining notations, modeling templates, analytical methods, and possibly other
useful operations on models of the model kind. 
A viewpoint can be instantiated for a domain at
hand, resulting in an \textit{architectural view}. 
An architectural view comprises architectural models that are developed using the conventions and methods established by its associated viewpoint. An architectural model may participate in more than one view. 
Although viewpoints have become increasingly popular for describing software architectures, their adoption in the domain of self-adaptive systems is rather limited.

\cite{6337733} proposed a variability viewpoint for constructing views of enterprise software systems. The viewpoint comprises several complementary models that address detailed variability concerns. The viewpoint facilitates the representation and analysis of variability (i.e., the ability of software to be adapted for a specific context to enable multiple deployments of a software system) in the architecture of an enterprise software system. 
\cite{10.1007/978-3-030-29983-5_8} documented an architecture viewpoint for continuous adaptation management in collective intelligence systems (CIS). The viewpoint frames concerns of stakeholders with an interest in handling CIS-specific adaptation across the entire system’s life cycle and includes a set of four model kinds for identifying, designing, and realizing adaptation in CIS key elements. The viewpoint provides support for software architects to deal with self-adaptive systems in collective intelligence systems. 

In summary, architectural viewpoints and views offer the means to document and analyze software designs from the perspective of stakeholder concerns. Leveraging a reference model (pointed out above), this chapter aims to present an architectural viewpoint for the decision-making component of self-adaptive systems. This viewpoint should consider the estimated benefit, cost, and risk of adaptation options in the decision-making process of self-adaptive systems.

\bibliographystyle{apalike}
\bibliography{bibliography}

\end{document}